\documentclass{aa}
\usepackage{graphicx}
\usepackage{txfonts}
\usepackage{natbib}
\bibpunct{(}{)}{;}{a}{}{,} 

\newcommand{\teff}[1]{$T_{\rm eff}$}
\newcommand{\vsini}[1]{$v\cdot\sin(i)$}
\def\cm1{$\rm cm^{-1}$}
\def\kms{$\rm km\,s^{-1}$}
\def\DE{D\kern-0.75em \raisebox{1.0pt}{=}\ }

\def\Rv{\rule[-0.07in]{0.0pt}{15.0pt}}
\def\Ru{\rule{0.0pt}{10.0pt}}

\def\II{{\sc ii\ }}
\begin{document}

   \title{The puzzling spectrum of HD 94509}

   \subtitle{Sounding out the extremes of Be shell star spectral morphology}
   \author{C. R. Cowley
          \inst{1}
          \and
           N. Przybilla
          \inst{2}
          \and 
          S. Hubrig 
          \inst{3}
          }

   \institute{Department of Astronomy, University of Michigan,
              Ann Arbor, MI 48109-1042, USA
              \email{cowley@umich.edu}
          \and
            Institut f\"{u}r Astro- und Teilchenphysik,
            Technikerstr. 25/8, A-6020 Innsbruck, Austria
             \email{Norbert.Przybilla@uibk.ac.at}
         \and
             Leibnitz-Institut f\"{u}r Astrophysik,
             Potsdam (AIP), An der Sternwarte 16, 14482, Potsdam, Germany
             \email{shubrig@aip.de}
             }
             
   \date{Received; accepted }

 
  \abstract
   {The spectral features of HD 94509 are highly unusual, adding an extreme
   to the zoo of Be and shell stars.  The shell dominates the
   spectrum, showing lines typical for spectral types mid-A to early-F, while
   the presence of a late/mid B-type central star is indicated by
   photospheric hydrogen line wings and helium lines.
   Numerous metallic absorption lines have 
			broad wings but taper to narrow cores.  They cannot be fit by Voigt
			profiles.}
   {To describe and illustrate unusual spectral features of this star,
    and make rough calculations to estimate physical conditions and 
    abundances in the shell. Furthermore, the central star is
    characterized.}
   {We assume mean conditions for the shell.  An electron density 
   estimate is made from the Inglis-Teller formula.  Excitation temperatures
   and column densities for \ion{Fe}{i} and \ion{Fe}{ii} are derived from curves of
   growth.  The neutral H column density is estimated from high Paschen members.
   The column densities are
   compared with calculations made with the photoionization code {\sc Cloudy}.
   Atmospheric parameters of the central star are constrained
   employing non-LTE spectrum synthesis. 
   }
   {Overall chemical abundances are close to solar.
				Column densities of the dominant ions of several elements, as well as
			excitation temperatures and the mean electron density are 
			well accounted for by a simple model.  Several features, including the
			degree of ionization, are less well described.
}
{HD 94509 is a Be star with a stable shell, close to the
terminal-age main sequence.   
The dynamical state of the shell and the unusually
shaped, but symmetric line profiles, require a separate study. } 


   \maketitle
%

\section{Introduction}\footnote{Based in part on
			ESO programs 082.C-0831, 083.A-9013, and 084.C-0952}

The unusual nature of the spectrum of \object{HD 94509} was noted during
a search of the ESO archive for possible Herbig Ae/Be stars.
Simbad lists HD 94509 as a Be star, while \citet{houk1}
call it Bp SHELL, with the remark that
it has strong typical shell lines, with H$\beta$ somewhat filled in.
The star appears among a number of surveys for young or unusual
objects, e.g. \citet{the} and \citet{reed}, that give 
little spectroscopic description beyond a basic classification.  

HD 94509 is a shell star by the modern criteria 
\citep{Hanuschik, riv}, which
required the central absorption of the emission lines to
be below the level of the underlying stellar continuum.
The typical picture of a shell star discussed
in these papers was with a flattened shell close to the
line of sight or nearly 90$^\circ$ to the plane of the
sky.   

\citet{wadeetal} included HD 94509 among their survey of
possible Herbig AeBe stars for magnetic fields,  but their paper
does not comment on other characteristics of the spectrum.
It appears in the catalogue of \citet{dewinter} as a Herbig Be
star.   However, the raw infrared colors 
($J-H$\,=\,+0.054, and $H-K$\,=\,$-$0.061) place HD 94509
within the group of classical Be stars, and well displaced from the
region of Herbig Ae/Be stars delineated by \citet{Jbox}.
Reddening corrections would only make the displacement larger.
But dilution indicators (see below) show that HD 94509 
is a shell star rather than a classical Be star.

We could find no detailed study of the spectrum, though
\citet[henceforth, CL]{Corpor} discuss HD 94509 in a note
on individual sources in their survey of binaries among Herbig
Ae/Be stars.  They noted \ion{He}{i}\,$\lambda$6678 as well as neutral lines
of \ion{Ca}{i}\ and \ion{Fe}{i}, and called the spectrum composite.  We confirm
the underlying \ion{He}{i} line, but see no evidence of a  
metallic-line spectrum from a secondary star.  
CL did not discuss the metallic-line profiles 
we find so remarkable.  

\section{A puzzling spectrum}

\subsection{Observations}

 The study is based primarily on UVES archive spectra of 
	HD 94509 obtained with the dichroic beam splitter
 DIC-2 in blue and red arms on 16 March 2009. The high-resolution
 spectrograph UVES \citep[Ultraviolet and Visual
 Echelle Spectrograph,][]{Dekkeretal00} is installed at the
 VLT-8m (ESO/Paranal, Chile). Our wavelength coverage is
 3750--4985, and 5682--9460\,{\AA}.
 The resolving power is $R\,=\,\lambda/\Delta\lambda\,\approx\,40\,000$, due
 to the use of a wide slit.  The signal-to-noise ($S$/$N$) varies but was 
 typically about 330 in the blue and 250 in the visual.

\begin{figure*}
\centering
\includegraphics[width=.84\linewidth]{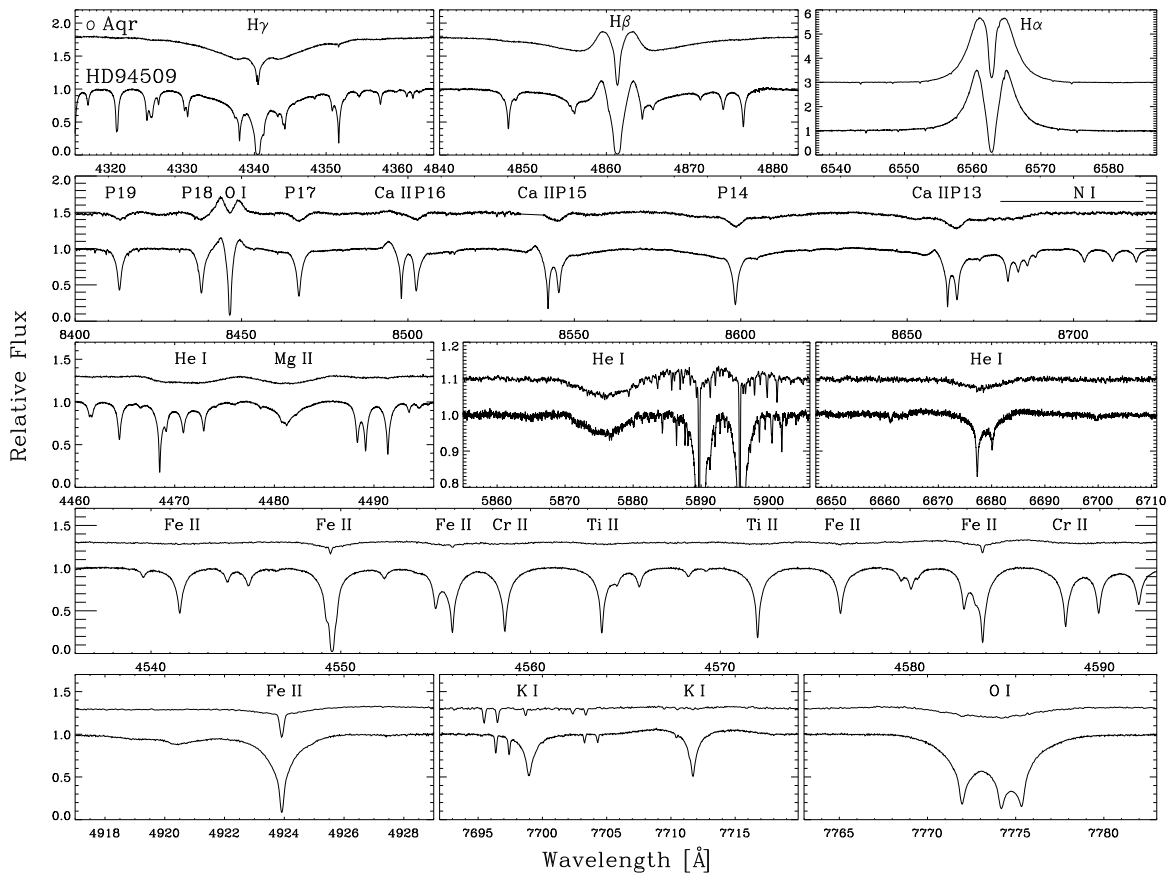}
\caption{Comparison of the spectrum of 
HD 94509 (lower curve) with a FEROS spectrum of $o$\,Aqr 
(B7-B8 IV-V, upper curve), 
a prototype Be shell star, for a selection of strategic lines. The most
prominent features are identified. Both spectra were
rv-corrected to the laboratory restframe.}
\label{fig:shellstars}
\end{figure*}

 We also used an ESO archive FEROS spectrum ($R$\,=\,48\,000, $S/N$\,$\approx$\,120) 
	obtained on 7 June 2009,
 covering the wavelength range from
	3666 to 8700\,{\AA}. FEROS \citep[Fiber-fed
 Extended Range Optical Spectrograph,][]{Kauferetal99} is a
 high-resolution, Échelle spectrograph operated at
 ESO in La Silla, Chile and is installed at the MPIA 2.2m telescope.
 Both UVES and FEROS spectra were examined for spectral variations. 
	The FEROS spectrum was also useful for features (e.g. \ion{Ba}{ii} $\lambda$4934)
 that were in UVES order gaps.
 Additionaly, an X-Shooter spectrum of HD 94509 from 5 February 2010 was
 retrieved from the ESO archive. X-Shooter \citep{Vernetetal11} is installed at VLT-8m
 (ESO/Paranal, Chile) and is the first of the second
 generation instruments at VLT.  The resolution and noise level of this
	spectrum was considerably lower than those of the UVES and FEROS spectra.  
	We only used the X-Shooter data to look for spectral variations.



Figure~\ref{fig:shellstars} shows a comparison of the spectrum of 
HD 94509 with that of $o$\,Aqr (\object{HD 209409}, ~B7-B8 V-IV), 
a prototype Be shell star, 
for a selection of strategic lines. While some similarities in the
Balmer and \ion{He}{i} lines are seen, the richness of the
metallic-line spectrum and the peculiar line profile shapes stand out,
and -- to our knowledge -- are unreported for any other star so far.
Note that $o$\,Aqr shows sharp line cores originating from the shell
in a few of the strongest \ion{Fe}{ii} lines, e.g. $\lambda$4924.

   \begin{figure}
   \centering
   \includegraphics[width=.87\linewidth]{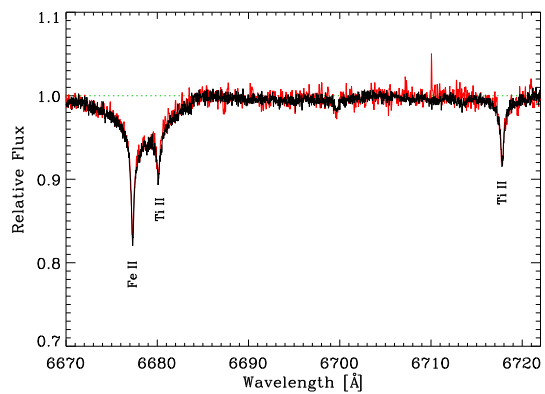}
      \caption{UVES (black) and FEROS (gray/red online) spectra in the region
      of CL's Fig.~20.  Three absorption features are identified as
      \ion{Fe}{ii} ($\lambda$6677.30), \ion{Ti}{ii} ($\lambda$6680.13), and 
      \ion{Ti}{ii} ($\lambda$6717.79).
                             }
         \label{fig:plcl}
   \end{figure} 

CL discuss three absorption lines in the neighborhood of the broad \ion{He}{i}
line $\lambda$6678, which they tentatively identify as \ion{Fe}{i} $\lambda$6677.989,
\ion{Ca}{i} $\lambda$6690.628, and \ion{Ca}{i} $\lambda$6717.681, with velocity shifts
of $-$45, $-$35, and $-$7 \kms.  Their Fig. 20 shows wavelength shifts at three
epochs, best seen in the strongest feature.
The upper part of CL's Fig.~20 closely resembles our spectra, shown in
detail in Fig.~\ref{fig:plcl}.  The broad \ion{He}{i} absorption is apparent.  
We can plausibly
identify the sharper absorptions as lines of \ion{Fe}{ii} and \ion{Ti}{ii} 
with the
radial velocity of the other metallic lines of the shell.
Interestingly, $\lambda$6677 of \ion{Fe}{ii} does not appear in the current 
 National Institute of Standards and Technology 
(NIST)\footnote{http://www.nist.gov/pml/data/asd.cfm} list,
though it is in the Multiplet Tables, and in
the Vienna Atomic Line Database 
(VALD)\footnote{http://vald.astro.univie.ac.at/$\sim$vald/php/vald.php}.  
G.~Nave of NIST (private communication) 
suggests it is masked in her spectra by an argon line.  
Our material 
does not indicate the presence of another star, or another shell with a relative 
radial velocity shift.  We can, of course, not exclude a stage of variability
at an epoch not covered by our spectra.

\subsection{The metallic lines\label{sec:metlins}}

\begin{figure}
   \centering
   \includegraphics[width=.87\linewidth]{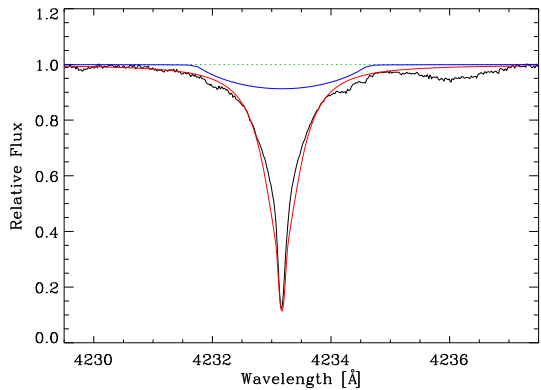}
      \caption{The strong \ion{Fe}{ii} line $\lambda$4233,
      illustrating the unusual metallic-line profiles.  The
      gray (red online) synthesis is based on a
      $T_\mathrm{eff}$\,=\,10000\,K, 
      $\log g$\,=\,2.5 non-LTE spectrum synthesis model with 10\,$\times$ solar abundances, but with the
      Stark broadening increased by a factor of 6\,$\times$\,$10^3$!
      The dark gray (blue online) rotational profile used 
      a 10\,$\times$\,solar iron abundance, with
      $V\cdot\sin(i)$\,=\,100\,\kms~and standard Stark broadening.}
         \label{fig:4233_DampRot}
   \end{figure}

Figure \ref{fig:4233_DampRot} is typical of the stronger 
metallic-line profiles.
A rough fit to the observed line is possible with a 10\,$\times$\,solar abundance
and a standard model atmosphere  see
Sect.~\ref{sect:characterization} for details on the modelling. 
However, the wings require an impossibly
large enhancement of the damping, by 6\,$\times$\,$10^3$. The figure also shows a calculated
profile based on a $V\cdot\sin(i)$ of 100\,\kms and standard Stark broadening\footnote{
Note that the effective temperature $T_\mathrm{eff}$ and surface gravity $\log
g$ of the model are chosen here for illustrative purposes only -- the line
is {\em not} formed in the atmosphere of the central star, but in the
shell.}. 

The stronger metallic-line profiles of HD 94509 gracefully taper to
cores as sharp as the instrumental resolution allows.   The profiles
merge with the continuum in what resemble Lorentz wings.  
The profiles are very symmetrical, with little or no
indication of a wind, or velocity asymmetry within the zone in which
they are formed.

   \begin{figure}
   \centering
   \includegraphics[width=.9\linewidth]{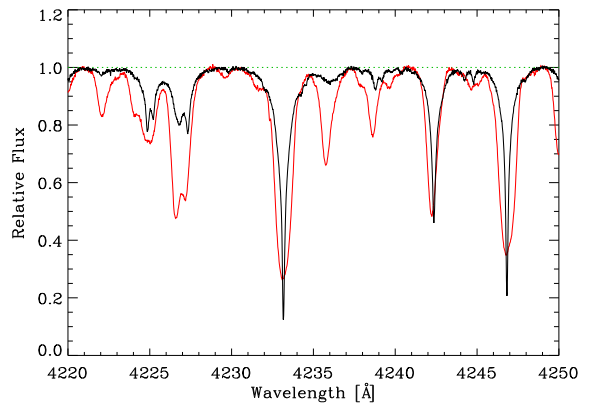}
      \caption
				{ To illustrate the richness of the metallic spectrum of
				HR 94509, we compare the region of the strong \ion{Fe}{ii} line 
				$\lambda$4233, in HD 94509
      (black)
      and that of the F3 Ia supergiant \object{HD 74180} 
						(gray, red online).  The somewhat
      rectangular shape of the profiles with rounded bottoms, 
      of the latter star
      may be rather closely matched by a model with appropriate
      microturbulent and rotational broadening.}
         \label{fig:4233}
   \end{figure}

A broader view of the $\lambda4233$-region is shown in 
Fig.~\ref{fig:4233}.  The richness of the HD 94509 spectrum is
comparable to that of the F3 Ia supergiant HD 74180,
and not common among shell stars, cf.~Fig.~\ref{fig:shellstars}.
Note we do not imply that HD 94509 is a composite containing
such a star.  Dilution indicators, discussed below (Section~\ref{sec:dilution}), 
show that the
metallic-line spectrum of HD 94509 is not photospheric.  
But a few
stable shell stars \citep{Gulliver} may have similar spectra.
\citet{merrill} shows a microphotometer tracing for \object{HD 193182},
whose spectrum is surely rich in metal lines.  We have been
unable to find a digital, high-resolution spectrum that would
allow us to make a close comparison of the line shapes. 

A few of the strong metallic lines show broad emission shoulders as 
well as relatively narrow, deep absorption.  
An example is 
the \ion{Fe}{ii} emission, $\lambda$6516: 
a$^6$S$_{2.5}$--z$^6$D$^o_{3.5}$ in 
Multiplet 40 \citep{moore}.  Multiplet 42 of \ion{Fe}{ii}
($\lambda\lambda$4924, 5018, and 5169) is commonly
in emission in many peculiar spectra (Be, Herbig Ae/Be, T Tauri).  It is
is also a sextet to sextet transition involving the same lower
(a$^6$S) term. Interestingly, Multiplet 42 is at most only
weakly in emission in HD 94509.
   
   \begin{figure}
   \centering
   \includegraphics[width=5.5cm,angle=-90]{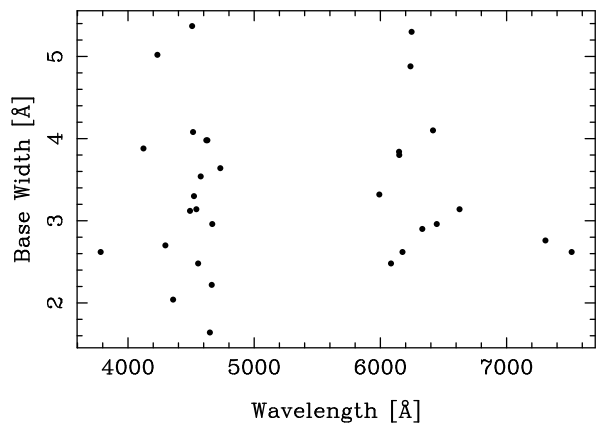}
      \caption{Base, or full width of 31 \ion{Fe}{ii} absorption 
               lines. An average half width for these lines corresponds to
               a velocity of some 100\,\kms. No correlation with wavelength is seen.
															 See text for uncertainties in measurements}
         \label{fig:plwid}
   \end{figure}

  \begin{figure}
   \centering
   \includegraphics[width=5.5cm,angle=-90]{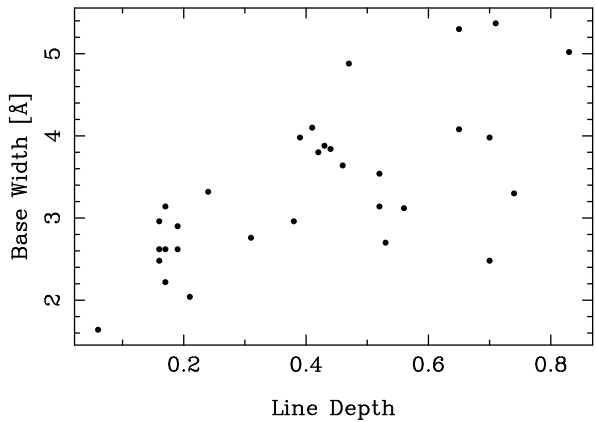}
      \caption{The base, or full width of \ion{Fe}{ii} absorption 
	lines shows a definite correlation with the line
	depths. For 31 points, the Pearson correlation coefficient
	is 0.674; the significance, or probability of no correlation
	is $3\times 10^{-5}$.}
         \label{fig:pwidd}
   \end{figure}

In this paper, we will not attempt to account for the 
overall line shapes, especially the graceful tapering
toward the cores of the stronger metallic lines.  
Figure~\ref{fig:4233_DampRot} shows that velocities
of the order of 100 \kms are required to account for the line
wings.  If the Doppler effect were responsible for the width
of these lines, a correlation of the width with wavelength
would be expected.  However, such a correlation is not obvious in 
Fig.~\ref{fig:plwid}.

Measurements of the base widths of absorption lines are most 
uncertain, because a qualitative judgement must be made of 
the position where the line merges with the continuum.  The
difficulty is compounded by the presence of blending.
Remeasurements of a sample of 10 randomly chosen Fe \II
lines shows an average difference of 12\%, with an extreme
value of 23\%.  
In 
spite of the uncertainty of these base widths, we do find 
a significant correlation of the base width with the line depth,
as shown in Fig.~\ref{fig:pwidd}.

      
We offer no explanation of the unusual line profiles
or the correlation of Fig.~\ref{fig:pwidd}.
Widths due to turbulent macroscopic motions would require
Mach numbers of 8 or more, and seem unlikely.  Rotation
of the central star, or Keplerian revolution of a disk  
could readily produce widths comparable to those measured.
It is puzzling that we cannot establish a clear correlation
of the line widths with wavelength.

Possibly calculations such as those
performed by \citet{HumDac}, but for absorption 
lines, could account for the line shapes.   We must leave 
this for future work.

 \subsection{The hydrogen lines}

The Balmer lines are highly symmetrical.  Both H$\alpha$ and
and H$\beta$ have double emission peaks of equal intensity, see
Fig.~\ref{fig:shellstars}.  The central absorption cores
yield velocities that closely agree with the metallic-line absorption.
The central reversals of both H$\alpha$ and H$\beta$ are very deep,
while the cores of H$\gamma$, through H$\epsilon$ are virtually black.
Our spectra extend from H$\alpha$ to H$\iota$ or H10, and none of
the core wavelengths deviate from the laboratory positions by more
than 0.04~\AA.  For such broad profiles, this may easily be attributed
to measuring error.  The mean deviation for the 9 lower Balmer cores
from that expected from the metallic cores is
0.0033~\AA.  The symmetry of both the hydrogen and metallic absorption
cores is indicative of ``negligible line-of-sight velocities'' in the
absorbing shell \citep[cf.][]{riv}.

   \begin{figure}
   \centering
   \includegraphics[width=.87\linewidth]{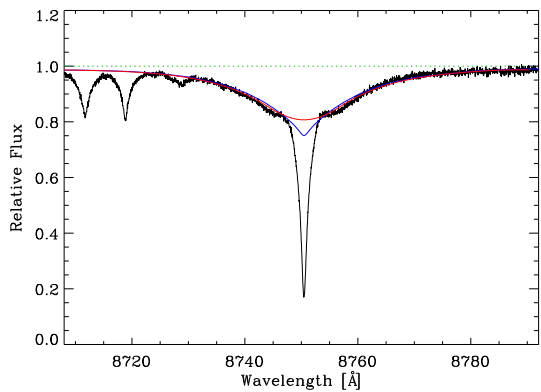}
      \caption{The Paschen line, P12, shows the sharp core 
      of the shell and the broad-line morphology of the central source.
      Overlaid are rotational profiles for
      0 (dark gray, or blue) and 260\,\kms (light gray, or red).
      Two \ion{N}{i} lines are seen to the violet of P12. See the 
      text for details.}
         \label{fig:P12}
   \end{figure}
   
   \begin{figure}
   \centering
   \includegraphics[width=.87\linewidth]{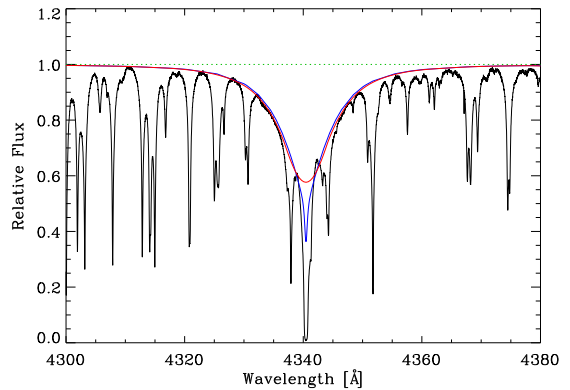}
      \caption{Like Fig.~\ref{fig:P12}, but for H$\gamma$. The black
      core of the line is indicative for the shell covering the entire
      stellar disk.}
         \label{fig:Hgamma}
   \end{figure}

The non-LTE calculations of P12 shown in Fig.~\ref{fig:P12} 
were made with a 14000\,K,
$\log g$\,=\,3.5 model,
assuming an additional continuum equal to 
a factor of 0.2 times the stellar continuum. Theoretical plots are
for $V\cdot\sin(i)$\,=\,0 and 260\,\kms. As the Stark-broadened wings of P12 are
of stellar origin (only the sharp line core originates in the shell),
this may be viewed in support of the atmospheric parameters chosen for
the central star (Sect.~\ref{sec:themod}). The same applies to other
hydrogen lines, also the Balmer lines from H$\gamma$ (see
Fig.~\ref{fig:Hgamma}) to the higher series members. 
A continuum contribution of 0.12 times the central star's photospheric 
continuum was adopted for H$\gamma$, meeting the expectation of a
decreasing ratio towards the blue. 
Note, however, that the true amount 
of dilution of the underlying spectrum is uncertain.

\subsection{The helium lines}
\begin{figure}
   \centering
   \includegraphics[width=.87\linewidth]{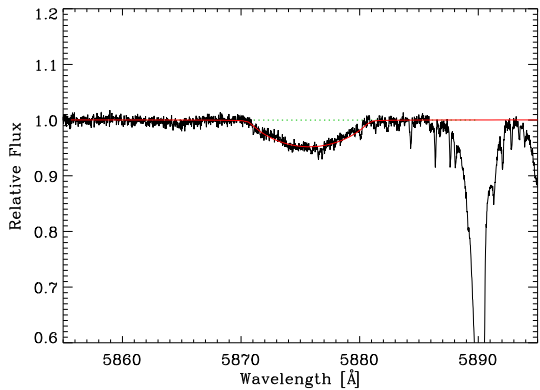}
      \caption{The \ion{He}{i} $\lambda$5875 line, which originates
      in the atmosphere of the central star, with no contribution from
      the shell. A rotational profile for
      260\,\kms~(light gray, or red) fits the observation. See the
      text for details.}
         \label{fig:He5875}
   \end{figure}

The \ion{He}{i} lines are broad and shallow, unlike the metal lines. 
They can be matched by rotationally-broadend 
line profiles, except for cases with metal line blends (see e.g.~Fig.~\ref{fig:plcl}).
This is indication for a stellar atmospheric origin of these lines.
They can be fit using atmospheric parameters as adopted for
the central star (Sect.~\ref{sec:themod}), and a solar helium abundance. Figure~\ref{fig:He5875}
shows the example of the line \ion{He}{i} $\lambda$5875, assuming an
additional continuum from the shell equal to a factor of 0.15 times 
the stellar continuum, intermediate between the two values adopted for
the hydrogen Balmer and Paschen line above. 
A rotational velocity of $\sim$260\,\kms~is indicated, in consistency with 
the typical picture of shell star, where the metallic absorption is
superimposed on the spectrum of a rapidly rotating central star.

\begin{table}
\small
\caption{Apparent rotational velocities (half widths)
in \kms.  Emission lines marked with an asterisk.  An
average is given for the \ion{Ca}{ii} infrared triplet.\label{tab:vsini}}             
\centering                          
\begin{tabular}{c c c }        
\hline\hline                 
{\Rv}Spectrum & Wavelength (\AA) & $V\cdot\sin(i)$  \\    
\hline                        
{\Ru}   \ion{He}{i}      & 5876 & 255   \\
   \ion{He}{i}      & 6678 & 299   \\
   \ion{Si}{ii}   &6347, 6371 & 245 \\
   \ion{Ca}{ii} (IRT)$^*$&8500-8660&225 \\
   Ave.~31 \ion{Fe}{ii}  &3783-7516 & 99$\pm 33$sd   \\
   \ion{Fe}{ii}$^*$&6516 & 214  \\
   \ion{K}{i}     & 7699 & 114   \\ 
\hline                                   
\end{tabular}
\tablefoot{$^*$ The width measurement is of emission shoulders.}
\end{table}

\subsection{Additional remarks}

Estimates of the putative $V\cdot\sin(i)$ may be made from the (half) 
base widths of both additional emission and absorption
lines.  Uncertainties in these values depend on the line width
measurements, whose accuracies are discussed in Section~\ref{sec:metlins}.
  
While the metallic spectrum is predominantly absorption, 
a few lines show ``emission shoulders''.  
Table~\ref{tab:vsini} summarizes the measurements, yielding the
following picture. The lines originating from the highest excitation potentials (i.e.~the
\ion{He}{i} lines) indicate the highest apparent rotational
velocities, of about 250--300~\kms. This should correspond to the
rotational velocity of the central star, which is in line with 
a typical $V\cdot\sin(i)$ that ranges from 200 to 450\,\kms~for the
central component of shell stars \citep{riv}. Slightly lower
velocities are obtained from the emission shoulders of spectral lines, 
which arise from the faster-spinning inner parts of the disk. 
Then, even lower rotational velocities are found for the metallic absorption lines,
which involve low-excitation levels and indicate the formation in the
slower-rotating outer parts of the disk. 

Finally, we note that several broad and shallow features are seen in the spectrum.
These coincide in wavelength with several of the typically most
pronounced diffuse interstellar bands (DIBs). In particular
$\lambda\lambda$5780.45, 6283.86 and 6613.62 are present, but shifted by
$\sim +$17\,\kms~relative to the intrinsic spectrum of HD 94509.
This is 
is consistent with the findings from the sodium D and calcium H and K
lines (see Sect.~\ref{sec:abunds}).



    \begin{figure}
   \centering
   \includegraphics[width=.99\linewidth]{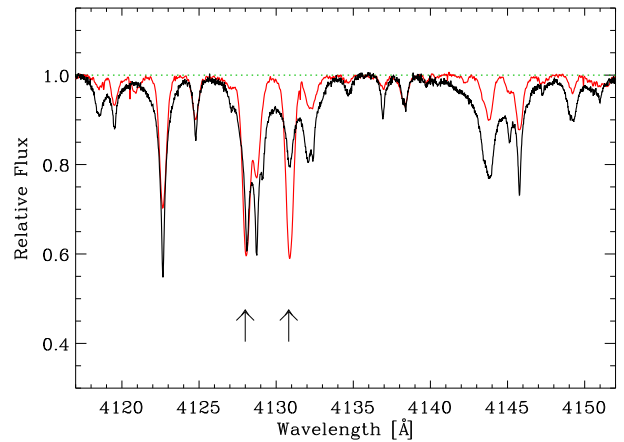} 
      \caption{Dilution in the shell of HD 94509.  The \ion{Si}{ii} lines 
						$\lambda\lambda$4128 
      and 4130 (arrows) are stronger in the A2 Iae 
      star Deneb (grey, red online) than in HD 94509 (black) because of dilution.
      The absorption at $\lambda$4128.12 in HD 94509 is largely due to
					 $\lambda$4128.13 of \ion{Mn}{ii}.      
	 \label{fig:Si2}}
   \end{figure}

\section{Characterization of the central star\label{sect:characterization}}
While the overall spectrum is dominated by shell absorption and
emission, a small
number of features are of photospheric origin, as indicated in the last
section: the wings of several hydrogen lines and \ion{He}{i}
lines. Photospheric metal lines unaffected by disk absorption cannot
be identified unambiguously.
These lines do not provide sufficient information for a fully-fledged 
quantitative analysis, but allow an estimate of the atmospheric parameters. 
A $T_\mathrm{eff}$\,=\,14\,000\,K and $\log g$\,=\,3.5, solar helium
abundance and $V\cdot\sin(i)$\,$\approx$\,260\,\kms~seems a proper
choice, but cannot be considered precise. The presence of
\ion{He}{i} lines set a lower boundary of the temperature to
$\sim$11\,000\,K. However, this would require an enhanced helium abundance,
which is unlikely in view of evolution models for rotating stars 
that do not predict a noticeable increase of the surface helium
values even at critical rotation in the relevant mass range 
\citep{Brottetal11,Georgyetal13}.
The wings of the hydrogen lines provide additional information
on the effective temperature, but are subject to temperature-gravity
degeneracy. Calculations with the ionization code CLOUDY 
(see Section~\ref{sec:cloudy}) show that the H$\alpha$ emission strength
would be too large for a higher temperature than the one adopted. 

The adopted gravity value implies that the star is 
in an advanced stage of the
main-sequence evolution, consistent with theoretical considerations
\citep[e.g.][]{Granadaetal13}. Note that the light from the central star
-- as a consequence of the fast rotation and the shadowing by the disk
seen equator-on (i.e. $i$\,$\sim$\,90$^{\rm o}$) -- is dominated by the hotter and slower-rotating
polar caps. Consequently, our above-mentioned parameters are likely 
upper values for the non-uniform $T_\mathrm{eff}$- and $\log g$-
distribution over the rotationally-flattened stellar surface, while the  
rotational velocity is a lower limit 
\citep[e.g.][]{Townsendetal04,Frematetal05}. The metallicity of the star
seems close to solar, or compatible with the present-day cosmic
abundance standard \citep{NiPr12}, as indicated by the analysis of the
shell (see Sect.~\ref{sec:cloudy}).

We employed a hybrid non-LTE approach for our stellar photospheric 
line-formation calculations, which has been successfully used before 
in a similar temperature regime to the one studied here
\citep{Przybillaetal06,NiPr07}. Plane-parallel, hydrostatic and
homogeneous LTE model
atmospheres were computed with the code {\sc Atlas9} \citep{Kurucz}.
Then, non-LTE line-formation computations were performed
with the codes {\sc Detail} and {\sc Surface} \citep[both updated
and improved]{Giddings81,BuGi85}. {\sc Detail}
calculates atomic-level populations
by solving the coupled radiative transfer and statistical equilibrium
equations, and {\sc Surface} computes the formal solution
using realistic line-broadening functions. We employed  
model atoms of \citet{PrBu04} for hydrogen, of \citet{Przybilla05} for
He, and of \citet{Becker98} for \ion{Fe}{ii}.

We note that the model calculations are intended to {\em illustrate} the
plausibility of our parameter choices, but are well aware that this
approach is suited to address the complex phenomenology of a Be
(shell) star only coarsly. We also note that similar approaches are
nevertheless employed for quantitative Be star analyses
\citep[e.g.][]{LeLe04,Dunstalletal11}.


\section{Physical conditions in the shell\label{sec:shell}}
\subsection{Dilution indicators\label{sec:dilution}}

A general characteristic of shell stars is the relative enhancement of
second spectra (from ions) relative to first (from neutrals), 
when compared with the spectra
of dwarfs or normal giants.  This may be attributed to the low
electron pressure in the shell, which will be discussed in more detail
below (cf.~Sect.~\ref{sec:IT}).  A second characteristic, discussed in 
numerous papers and reviews by Struve \citep[e.g.][]{Struve}, is the weakness of 
certain lines which he called ``dilution indicators''.  Specifically,
Struve called attention to lines of \ion{Si}{ii} $\lambda\lambda$3853-3862,
4128-4130, and \ion{Mg}{ii} 4481.  He noted that the lower levels of the
dilution indicators were connected to the ground level by strong 
permitted transitions.  Most absorption lines of \ion{Fe}{ii} and
\ion{Ti}{ii} arise from levels of the same parity as the ground term, and are 
thus metastable.    

We illustrate this in Fig.~\ref{fig:Si2},
where lines in HD 94509 and the supergiant \object{Deneb} are shown.  
The latter spectrum is the one described by \citet{SchPr08}.
Generally, the metallic lines in HD~94509 are significantly stronger than in Deneb,
but this is not the case for Struve's dilution indicators.

\subsection{Electron density\label{sec:IT}}
A depth-dependent model of the shell of HD 94509 is beyond the scope 
of the present work.  We attempt to determine mean conditions for
the shell as a whole, making use of techniques common for normal stellar
atmospheres prior to the use of numerical modeling.  

\begin{table}
\small
\caption{Electron densities $\log (N_e)$}             
\label{tab:ITres}      
\centering                          
\begin{tabular}{c c c}        
\hline\hline                 
{\Rv}constant &$n=41$ & $n=42$  \\    
\hline                        
{\Ru}   
   23.26 & 11.16 & 11.08 \\      
   23.49 & 11.40 & 11.32 \\
\hline                                   
\end{tabular}
\end{table}

Our high-resolution spectra do not include the Balmer limit, 
but the Paschen limit is 
clearly seen.  The line P41 was measured, but P42 was not, and an 
inspection of the spectrum reveals no obvious feature between P41
and three moderate \ion{N}{i} lines.  A well-established interpretation of
the termination of a one-electron line series is that overlapping
Stark wings merge so individual features are no longer seen
\citep[henceforth, IT]{it}.  A simple
derivation of the  formula may be found in \citet{urOpus},
where it is assumed that protons and electrons contribute
equally to the line merging, unlike IT.  If the last Paschen line is P$n$,
then
\begin{equation}
\log(N_e) = \log(N_p) = 23.26 - 7.5 \log(n)\,,
\label{eq:IT}
\end{equation}
where $N_e$ and $N_p$ are the electron and proton number densities,
respectively. \citet{varsh} uses the constant 23.49 rather than 23.26.
Let us put $n = 41$ and 42 into Eq.~\ref{eq:IT}.  Results are shown in
Table~\ref{tab:ITres}.  We conclude that a mean logarithmic 
electron density of the shell is 11.2$\pm$0.5.

\subsection{The shell temperature\label{sec:shellT}}
\subsubsection{\ion{Fe}{ii}\label{sec:Fe2}}

We know from the dilution indicators that there can be no single 
temperature that will describe the excitation and ionization within
the shell.  Nevertheless, the ionization and excitation may be 
approximately estimated from measurements of the line strengths.
These estimates will be subject to clarification by
models of the shell.  We have worked primarily with the
second spectra of metallic lines: \ion{Sc}{ii}, \ion{Ti}{ii},
\ion{Cr}{ii}, \ion{Fe}{i}, 
\ion{Fe}{ii}, and \ion{Ni}{ii}.
The \ion{Fe}{ii} spectrum is by far best suited for our purposes, both
because of its richness and the detail of its treatment by the
versatile program {\sc Cloudy} \citep[c13.03,][cf.~Sect.~\ref{sec:themod}]{Fer}.

\begin{figure}
   \centering
   \includegraphics[width=5cm,angle=-90]{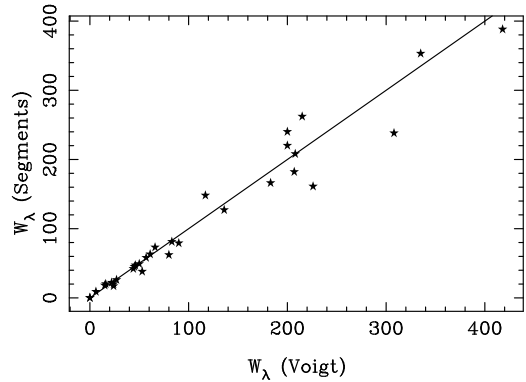}
      \caption{Comparison of $W_\lambda$'s (m{\AA}) measured using segmented 
      profile fits and those using fits to a Voigt profile.
					 the average absolute value of the differences in the
						plotted equivalent widths is 14 m\AA} 
         \label{fig:compare}
   \end{figure}

The unusual shape of the \ion{Fe}{ii} (and other metallic line) profiles
means that measurements of equivalent widths $W_\lambda$ require special care.  
Our general approach
in abundance work is to fit metal lines with Voigt profiles.  However,
the $a = \gamma/(2\Delta\lambda_D)$ is usually small, typically 0.01 to 0.1,
so the profiles are not far from Gaussian.  Such a low $a$-value is
completely unsuitable for HD~94509.  For the lines with profiles that
are not too deep, we could make tolerable fits using $a\,=\,5$, virtually
Lorentzian profiles.  Core and wings of the stronger (deeper) lines 
could not be fit by any Voigt profile. All of the profiles were also  
fit with small line segments, so for many lines we could compare the
results (Fig.~\ref{fig:compare}).  We adopted simple means for all pairs
of measurements.

\begin{figure}
   \centering
   \includegraphics[width=5cm,angle=-90]{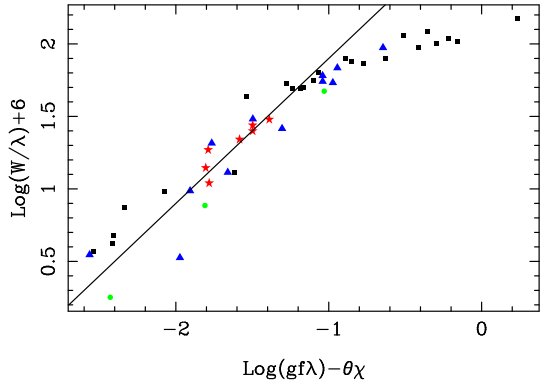}
      \caption{Excitation temperature from \ion{Fe}{ii} lines in the shell.
      Colors may be seen online.
      The (red) stars indicate lines with the highest lower excitation
      energies (6.2 to 7.3\,eV).  The lowest-excitation lines (2.6-2.9\,eV)
      are indicated by black squares.  The black line indicates a 45$^\circ$
      slope. This plot was made with $\theta\,=\,0.55$ or $T$\,=\,9164\,K
      with an uncertainty of $\sim$170\,K (see text).}
         \label{fig:plgcg}
   \end{figure}

One may estimate an overall excitation temperature by plotting 
$\log(W_\lambda/\lambda)$ vs. $\log(gf\lambda)- (5040/T)\cdot \chi$, where
  $\chi$ is the lower excitation potential of the absorption line.  One 
  adopts the temperature $T$ as the value that gives the minimum separation of
  lines with different $\chi$.  We illustrate this determination
  using reliable \ion{Fe}{ii} $gf$-values, from \citet[cf. also Adelman~2014]{melen}.
		Figure~\ref{fig:plgcg} shows the
  adopted fit, using $\theta\,=\,5040/T\,=\,0.55$, 
  or $T_{\rm exe}\,=\,9164\pm 170$\,K.  The uncertainty is derived from
		trial plots with $\theta = 0.55$ and 0.56, which were judged marginally 
		acceptable.
   
\subsubsection{\ion{Fe}{i} and \ion{Ti}{ii}}

Other atomic species with enough measurable (mostly 
unblended) lines to allow a credible
curve of growth are \ion{Fe}{i}  and \ion{Ti}{ii}.  Results are gathered in 
Table~\ref{tab:othercg}; uncertainty estimates, explained in \ref{sec:Fe2}, are
significantly larger for \ion{Ti}{ii} and \ion{Fe}{i}, which have
fewer lines than \ion{Fe}{ii}.
\begin{table}
\small
\caption{Excitation temperatures.  See text for uncertainty
estimates.}             
\label{tab:othercg}      
\centering                          
\begin{tabular}{c c c c c}        
\hline\hline                 
Spectrum &No. lines & $\chi$\,(eV) range&$T_{\rm ex}$\,(K)& Uncertainty (K) \\    
\hline                        
{\Ru}\ion{Ti}{ii}  & 26  &0.62 - 3.12 & 9600 & 600\\      
   \ion{Fe}{i}     & 29  &0.00 - 3.63 & 7750 & 650 \\
   \ion{Fe}{ii}    &46   &2.58 - 7.27 & 9164 & 170  \\
\hline                                   
\end{tabular}
\end{table}

  \subsection{Column densities \label{sec:NH}}
  
  		The column density of an absorber follows from the relation for {\it weak} 
		lines with equivalent widths~$W'$:
\begin{equation}
       W' = \frac{\pi e^2}{mc^2}\lambda^2 N_n f_{nm}.
       \label{eq:wprimed}
 \end{equation}      
\noindent Here, $N_n$ is the number (per cm$^{-2}$, column density) 
in the lower energy level, in the absorbing shell.  We have assumed the metal
line absorption arises entirely in the shell. In this context we note
that the central star has a spectral type in the late/mid-B range and
it is unlikely to find chemical peculiarities, as the fast rotation
inhibits the CP phenomenon because of
rotationally-driven hydrodynamical mixing. Therefore, one may expect
the only features of iron-group species present in the stellar spectrum 
to be weak \ion{Fe}{ii} lines. At most, these stellar lines contribute a few percent
to the measured equivalent widths, i.e. they are indeed negligible.
In fact the presence of strong unusually-shaped features of e.g.
\ion{Sc}{ii}, \ion{Ti}{ii}, \ion{Cr}{ii}, \ion{Fe}{i} and
\ion{Ni}{ii} makes the spectrum appear so unusual.

  \subsubsection{\ion{Fe}{ii}\label{sec:fe2}}
  One may estimate from Fig.~\ref{fig:plgcg} 
  that lines as strong as 250\,m{\AA}
  ($\log(W_\lambda/\lambda$\,+\,6\,=\,1.7,
  for 5000\,\AA) are still on the linear part of the curve of growth.  Let us 
  make the provisional assumption that iron is predominantly first ionized,
  neglecting \ion{Fe}{i} and \ion{Fe}{iii}.

Using the Boltzmann formula,
with $u(T)$ the partition function of Fe$^+$, and using logarithms, this
may be written to give the column density of all Fe$^+$ for each line.  With 
$\theta \equiv 5040/T$, and the lower excitation energy $\chi_n$, we have
\begin{equation}
\log[N(\mathrm{Fe^+})] = \log(W'_\lambda)/\lambda)+12.053-\log(gf\lambda)+\log(u(T))+\theta\chi_n.
\label{eq:logNH}
\end{equation} 

We work out the column densities for Fe$^+$ using Eq.~\ref{eq:logNH} for 
the \ion{Fe}{ii} lines shown in Tab.~\ref{tab:FeNH}.
The average value (of the logarithms) of the last column is
18.77$\pm$0.19.
\begin{table}
\small
\caption{Calculation of column densities for Fe$^+$, 
Dimensions are given for the arguments of the logarithmic
quantities.}
\label{tab:FeNH}
\centering
\begin{tabular}{c c c c c}\hline\hline
{\Rv}   $\lambda$& $\chi$   & $\log(gf)$ &  $\log[W_\lambda]$  &
$\log[N(\mathrm{Fe}^+$)]     \\
   \AA      &eV     &no units    &m\AA     &cm$^{-2}$  \\   \hline 
{\Ru}   4147.27  & 4.616 & $-$3.508&    0.88&     18.54 \\
   4413.60  & 2.676 &  $-$3.79&    1.76&     18.58  \\
   4601.40  & 2.891 &  $-$4.48&    1.34&     18.93  \\
   4648.94  & 2.583 &  $-$4.58&    1.54&     19.05  \\
   4720.15  & 3.197 &  $-$4.48&    1.22&     18.96  \\
   4833.20  & 2.657 &  $-$4.64&    1.31&     18.89  \\
   4840.00  & 2.676 &  $-$4.75&    1.25&     18.95  \\
   4893.82  & 2.828 &  $-$4.21&    1.67&     18.90  \\
   5813.67  & 5.571 &  $-$2.51&    1.65&     18.54  \\
   6084.11  & 3.199 &  $-$3.79&    2.10&     18.93  \\
   6103.51  & 6.217 &  $-$2.171&   1.93&     18.80  \\
   6317.40  & 6.223 &  $-$2.16&    1.84&     18.67  \\
   6627.22  & 7.274 &  $-$1.609&   2.09&     18.91  \\
   7301.57  & 3.892 &  $-$3.63&    1.85&     18.74  \\
   7515.81  & 3.903 &  $-$3.39&    1.99&     18.62  \\
   7841.40  & 3.903 &  $-$3.721&   1.42&     18.35  \\ \hline
   average  &{\Rv}  &       & & 18.77$\pm$0.19\,(sd) \\  \hline
   \end{tabular}
   \end{table}

\subsubsection{Neutral iron \label{sec:neutfe}}

While the \ion{Fe}{i} lines are less numerous than those of \ion{Fe}{ii},
a number of them may be measured, and analyzed in a similar manner.
Oscillator strengths for \ion{Fe}{i} are from
\citet{FuhrWiese}. Table~\ref{tab:Fe1NH} presents the results.
The average of the logarithms for the last column is 
15.41$\pm$0.35.

\begin{table}
\small
\caption{Calculation of column densities for neutral Fe
or Fe$^0$. 
The format is the same as for Table~\ref{tab:FeNH}.
\label{tab:Fe1NH}}
\centering
\begin{tabular}{c c c c c}\hline\hline
 {\Rv}  $\lambda$& $\chi$   & $\log(gf)$ &  $\log[W_\lambda]$  &
 $\log[N(\mathrm{Fe}^0$)]     \\
   \AA      & eV    &no units    &m\AA         & cm$^{-2}$ \\  \hline
{\Ru}3873.76 & 2.43&  $-$0.88&  1.72&  15.52 \\   
3983.96 & 2.73&  $-$1.02&  1.45&  15.54 \\   
4118.55 & 3.63&   0.22&  1.88&  15.20 \\   
4134.68 & 2.83&  $-$0.68&  1.53&  15.30 \\   
4175.64 & 2.84&  $-$0.83&  1.32&  15.24 \\   
4181.76 & 2.83&   0.37&  1.75&  14.46 \\  
4309.37 & 2.95&  $-$1.19&  1.83&  16.15 \\   
4442.34 & 2.20&  $-$1.26&  1.18&  15.11 \\   
4459.12 & 2.18&  $-$1.28&  1.48&  15.42 \\   
4466.55 & 2.83&  $-$0.60&  1.69&  15.32 \\   
4476.02 & 2.84&  $-$0.82&  1.52&  15.37  \\  
4494.56 & 2.20&  $-$1.14&  1.62&  15.43  \\  
8688.63 & 2.18&  $-$1.21&  1.96&  15.26  \\ \hline
average &{\Rv}&     &  &15.33$\pm$0.35\,(sd)  \\ \hline   
   \end{tabular}
   \end{table}

\subsubsection{\ion{Sc}{ii}, \ion{Ti}{ii}, \ion{Cr}{ii}, and
\ion{Ni}{ii}\label{sec:others}}

Column densities have been estimated for other species,
using the \ion{Fe}{ii} curve of growth to choose ``weak'''
lines or estimate saturation corrections.
Results are discussed in Sect.~\ref{sec:cloudy}
below.  These are uncertain by at least a factor of three 
because of the difficulty in estimating the contributions
from the wings, especially in the case where wings from
different lines overlap.  Non-LTE effects are also expected,
as can be seen from the more detailed results for \ion{Fe}{ii}
available from {\sc Cloudy} (cf.~Sect.~\ref{sec:cloudy}).  


A resume of observed and theoretical column densities may
be found in Sect.~\ref{sec:cloudy}. 
Relevant data are presented
in Table~\ref{tab:others} in Appendix~\ref{appendixB}.

\subsubsection{Hydrogen \label{sec:NHH}}
 
The highest Paschen lines (those that have no contribution from the
central star) may be used to make a crude estimate of
the column density $N_3$, 
of hydrogen in the $n = 3$ level.  Measurements of equivalent widths
were made using the segmented fits described above.    
With Eq.~\ref{eq:wprimed}, we
obtain values of $N_3$ that would apply {\it as if} 
the lines were
unsaturated.  In the ideal case, discussed in the classical text
of \citet{unsold}, we would expect values of $N_3$ to increase, 
as the lines become optically thin, and then to decrease as the lines
blend together, so their entire width is not measured.  Results are 
shown in Table~\ref{tab:N3H}.  The line P35 was blended and hence
not used.  Absorption oscillator strengths were calculated using 
exact expressions for the wave functions and relevant integrals.

\begin{table}
\small
\caption{Calculation of column densities 
$N_3$ (cm$^{-2}$)\, for $n=3$,
from Paschen lines, P$n$, where $n$ is in the first 
column.}  
\label{tab:N3H}
\centering
\begin{tabular}{r c c c}\hline\hline
{\Rv}$n$&$W_\lambda$(m{\AA})&$f$&  $\log(N_3)$     \\  \hline
{\Ru} 30& 368.&   2.034E-04 &  15.474  \\
 31& 354.&   1.840E-04 &  15.501  \\
 32& 259.&   1.670E-04 &  15.408  \\
 33& 210.&   1.521E-04 &  15.358  \\
 34& 186.&   1.388E-04 &  15.345  \\
 35&     &   1.271E-04 &          \\
 36& 160.&   1.167E-04 &  15.355  \\
 37& 138.&   1.075E-04 &  15.327  \\
 38& 116.&   9.901E-05 &  15.288  \\
 39&  90.&   9.150E-05 &  15.212  \\
 40&  89.&   8.474E-05 &  15.241  \\
 41&  73.&   7.863E-05 &  15.188  \\
 42&  61.&   7.309E-05 &  15.138  \\  \hline
 average&{\Rv} & & 15.32$\pm$0.11\,(sd)\\ \hline
   \end{tabular}
   \end{table}
   
The last column of Table~\ref{tab:N3H} decreases as $n$ increases,
as expected.  We do not see a decrease due to saturation for the
smaller $n$.  However, the variation 
from P30 to the last visible line, P42, is not large, so the uncertainty
is comparable to other approximations made herein.  We adopt
$\log(N_3)\,=\,15.3$.    



 \section{Qualitative chemistry of the circumstellar envelope or shell
	 \label{sec:abunds}}


Of the lighter elements, hydrogen is well represented by the 
features described previously.  We see broad absorption from \ion{He}{i}
$\lambda$5876 and 6678, and most of the bluer optical lines, 
but this is from the underlying stellar spectrum.  There is no sign of 
emission in the \ion{He}{i} features.  There is also not a hint
of absorption or emission at the position of the \ion{Li}{i} resonance
lines near $\lambda$6708 -- which is expected because any lithium
initially present would be transported to deeper layers of the stellar
envelope due to rotationally-induced mixing, and 
destroyed by nuclear reactions.   
This applies in a similar manner to beryllium and boron 
(relevant lines are located outside the available spectra). 
Absorption near $\lambda\lambda$9062 and 9112 could be due
to \ion{C}{i} in Multiplet 3. No \ion{C}{ii} lines were noted.
A number of absorptions due to \ion{N}{i} were identified in the near 
infrared. Several of these appear in Figs.~\ref{fig:shellstars} and \ref{fig:P12}.  The sharp
cores of the \ion{N}{i} lines are indicative of absorption
from the shell.  Neutral oxygen lines from the shell are
the strong infrared triplet $\lambda\lambda$7772, 7774, and 7775 and the line $\lambda$8446 (see
Fig.~\ref{fig:shellstars}).   

The D lines of sodium are strong.  Each line is double, with
the deeper component having the stellar radial velocity, with the 
strong, second component shifted to the red by 17.3\,\kms, likely
being of interstellar origin (an analogous situation is seen
in the calcium H and K lines).
Several lines of \ion{Mg}{i} are identified including the b-lines, which
have the broad wings and sharp cores of the shell.  The usually
strong \ion{Mg}{ii} $\lambda$4481 is severely weakened as noted 
earlier.
The \ion{Al}{i} resonance doublet is present, but weak.  No \ion{Si}{i} lines
were identified, but \ion{Si}{ii} lines are present in addition to 
the dilution indicators mentioned in Sect.~\ref{sec:dilution}.
The violet component of the \ion{K}{i} resonance pair, $\lambda$7665 is
obscured by water-vapor absorption, but $\lambda$7699 is unblended
and strong ($W_\lambda$\,$\approx$\,500\,m{\AA}), see
Fig.~\ref{fig:shellstars}.  The 3$d$ elements Ca-Ni are all well
represented, but heavier elements are weak or absent altogether,
similar to the situation in a supergiant spectrum at comparable
temperature.

An analysis by wavelength coincidence statistics (WCS) reveals meager 
evidence for the presence of neutron addition elements. Even the
\ion{Sr}{ii} resonance lines, $\lambda\lambda$4077 and 4215 are weak and blended.
Neither of the \ion{Ba}{ii} resonance lines, $\lambda\lambda$4554 and 4934 show
more than a {\it possibility} of absorption
near their positions.  This need not imply an underabundance of Ba or Sr
as we anticipate significant double ionization of these two elements.  
\ion{Zr}{ii} may be weakly present, but there is no evidence for \ion{Y}{ii}.

\begin{table}
\small
\caption{Characteristics of the {\sc Cloudy} model  which
fit the observations as summarized in Table~\ref{tab:compar}.
Keplerian velocities, $V_K$ are in \kms, for a 4.2\,$M_\odot$ star.}             
\label{tab:clmod}      
\centering                          
\begin{tabular}{l l}        
\hline\hline                 
{\Rv}Central star & $T_{\rm eff}= 14000$K, $\log g = 3.5$  \\
Stellar luminosity& $2.4\times 10^{36}$ erg s$^{-1}$  \\
Stellar radius &  4.2\,$R_\odot$  \\
Inner cloud radius& 14.4\,$R_\odot$; $V_K = 236$ \kms \\
Outer cloud radius & 45.5\,$R_\odot$; $V_K = 133$ \kms  \\
Cylinder semi-height & 5.0\,$R_\odot$ \\
Total hydrogen density& $1.58\cdot 10^{11}$\,cm$^{-3}$ (constant)  \\
Average electron density&$9.8\cdot 10^{10}$\,cm$^{-3}$ (variable)  \\
Molecules and grains & off  \\
 \hline                        
\end{tabular}
\end{table}

\section{A model using {\sc Cloudy}\label{sec:cloudy}}
\subsection{Details \label{sec:themod}}

It is clear that the conditions in the shell cannot be well described
by local thermodynamic equilibrium (LTE).  We have therefore
used the versatile photoionization code {\sc Cloudy} \citep{Fer}
to broaden our analysis.  Ideally, we would run {\sc cloudy} as a 
subroutine, and try to minimize deviations of its predictions against
our measurements.  Such an approach is beyond the scope of the present
work.  We have limited ourselves to trial and error comparisons of 
selected measurements or derived quantities, assuming default
solar abundances \citep{asplund,scota,scotb,grev14}.

The stellar background is from a \citet{Kurucz} {\sc Atlas9} 
model.  Details, and some relevant quantities are collected in 
Table~\ref{tab:clmod}.  Properties for the $T_{\rm eff}$\,=\,14\,000\,K
main sequence star were chosen with the help of data from
\citet{torres}.
We report results here for a
cylindrical geometry described in the documentation for
{\sc Cloudy} (Hazy-1, Sect.~9.6).  Similar results were obtained
with a spherical geometry with the same parameters apart from the
cylinder height.  

\begin{figure}
   \centering
   \includegraphics[width=5cm,angle=-90]{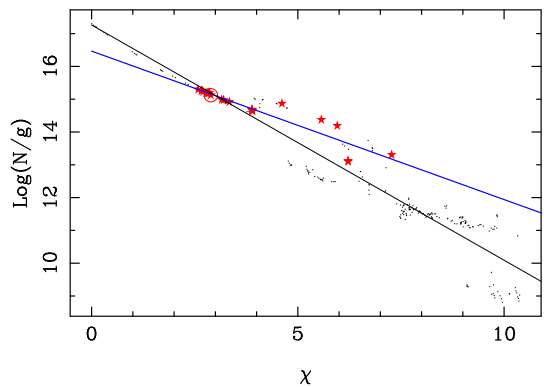}
      \caption{Theoretical ``excitation temperatures'' from column 
   densities for \ion{Fe}{ii} levels of the {\sc Cloudy} models.  The
			quantity $g$ is the statistical weight, $2J + 1$, of the levels.
			The larger symbols are for levels that correspond to the 
			lines used in Fig.~\ref{fig:plgcg}. The heavy line is a 
			least squares solution giving $T$\,=\,11134\,K.
			Smaller symbols are for 
			all 256 levels used in the model; the lighter line 
			corresponds to $T$\,=\,7026\,K.     
         \label{fig:pltex}}
   \end{figure}

It is possible to obtain column densities for any one of  
371 individual levels of Fe$^+$.  Our models used
256 levels, to save computing time.  We may obtain an
``excitation temperature'' from these column densities that
should be comparable to the values derived by the empirical
curves of growth from \ion{Ti}{ii}, \ion{Fe}{i}, and \ion{Fe}{ii}.  This is
shown in Fig.~\ref{fig:pltex}.

 Recall that the empirical results for \ion{Fe}{ii} and \ion{Ti}{ii} were 
	$\sim$9164\,K, while that for \ion{Fe}{i} was 7750\,K.  Theoretical and
	empirical results are summarized in Table~\ref{tab:compar}.
	The ``empirical'' entries for neutral and ionized hydrogen
	are described as semi-empirical and shown inside brackets.
	While the IT formula allows us to estimate a mean electron density
	(cm$^{-3}$), we can only infer a column density indirectly.
	The entry for H$^+$ in Table~\ref{tab:compar} comes from
	taking the product of the IT electron density and multiplying
	by the physical thickness of the {\it model}, to obtain a 
	column density.  We assume the column density of protons is 
	reasonably approximated by the electron column density thus
	obtained.
	
	We have an empirical estimate of the column density only for the
	$n\,=\,3$ level of neutral hydrogen.  We need a temperature to
	convert this to a column density for all levels of hydrogen,
	using the Boltzmann formula.
	There is no obvious empirical value to use; 
	indeed, because of the non-LTE conditions
	the meaning of any temperature so chosen is strictly limited
	to the Boltzmann formula and the relevant level populations.
	If we simply define a $T_{\rm ex}$ as that value giving the
	column densities $N_3$(H) and $N$(H), the adopted CLOUDY 
	{\it model} gives
	$5040/T_{\rm ex}$\,=\,0.69.  Recall the empirical values of 
	$\theta$ from \ion{Fe}{i} and \ion{Fe}{ii} were 0.65 and 0.55.  The entry
	for H$^0$ in Table~\ref{tab:compar} was obtained using
	$\theta$\,=\,0.69, and the empirical $\log(N_3$(H))\,=\,15.3.  
	
	 Let us assume (Hanuschik 1996) that Be and shell stars have
	``disks that are intrinsically identical.'' Then our basic
	 parameters from Table~\ref{tab:compar} should be comparable with
		those estimated for Be stars.  These stars display a variety of
		characteristics, summarized by Slettebak and Smith (2000).  The
		stellar types range from late O (X Per) to late B ($\beta$ CMi).
		The disks are typically 5 to 20 stellar radii, with temperatures
		of 10 000K, and electron densities of 10$^{10}$ to 10$^{13}$ 
		per cm.  Our values for $N_{\rm e}$, $T$, and radii agree
		well with these values. 
	
 \begin{table}
\small	
	\caption{Comparison of various empirical results with
	corresponding quantities from the model using solar
	abundances.
	Numerical values without units are logarithms of column
	densities (cm$^{-2}$).  Neutral species are indicated by a
	superscript `$0$'. Parenthesized numbers indicate the
	number of lines used.  Values in square brackets are
	semi-empirical (see text).\label{tab:compar}}
	\centering
 \begin{tabular}{l c || c}
 \hline\hline
 {\Rv}Species  &  Model       &Empirical  \\  \hline
 {\Ru}H$^0$& 23.30        & [22.8]  \\
 H$^+$    & 23.15             & [23.5] \\
 N$_3$(H) &15.85              & 15.3   \\
 $\log(N_e)$(cm$^{-3})$ &10.99&11.2  \\
 Sc$^+$   & 14.54             & 15.2(2)    \\
 Ti$^+$   & 16.43             & 16.8(5) \\
 Cr$^+$   & 17.11             & 17.2(4)  \\
 Fe$^0$& 14.25            & 15.4(13)  \\
 Fe$^+$   &18.98              & 18.8(16)  \\
 Ni$^+$   &17.74              & 17.6(4) \\  
	$T_{\rm ex}$\,(K)&11134        &9164   \\
	$W_\lambda$(H$\alpha$)\,(\AA)&28 & 24   \\ \hline
 \end{tabular}
	\end{table}

   \subsection{Comments \label{sec:results}}


	The dominant species, e.g. Fe$^+$ are predicted reasonably well
	by the model.  For column and particle densities 
	in Table~\ref{tab:compar}, apart
	from neutral iron, the rms deviations of the empirical 
	from the 9 model values is 0.33 dex, or a factor of 2.1.  
	Neutral iron is an order of magnitude
	or more stronger than the predictions.  
 If we include neutral iron, the rms deviation is 0.42 dex. 
	A cooler central star 
	could reduce this discrepancy, but the broad \ion{He}{i} absorptions
	would then be difficult to explain.  It does not help the 
	ionization discrepancy to move the cloud out by a factor of 10.
	Higher or lower overall densities were also tried, with 
	appropriate adjustments of the absorption path lengths.  
	
	We note that the treatment of complex atoms of the 
	iron group by {\sc Cloudy}, apart from \ion{Fe}{ii}, 
	does not involve multi-level
	models.  In view of this approximate treatment,  
	we do not consider the discrepancy with the \ion{Fe}{i}
	column density to be serious in the context of the present
	study.

\section{Conclusions}

HD 94509 is a shell star with an unusually stable shell, posing an
extreme in the zoo of Be star spectral morphology.  We have
not been able to account for the unusual shapes of the metallic
absorption lines.  However a simple model of the shell can predict
the total absorption from \ion{H}{i}, the electron density, 
and column densities of the dominant ions. Near-solar elemental
abundances are found. The lowest Balmer
lines have deep cores and symmetrical emissions.  The cores of
H$\gamma$ through H$\epsilon$ are virtually black, indicating the
shell must cover the stellar disk.

\begin{acknowledgements}
CRC thanks Howard Bond, Francis Fekel, and George Wallerstein for comments on 
samples of the HD 94509 spectrum.  Advice and help from J. Hernandez and numerous 
Michigan colleagues is gratefully acknowledged.  
We thank Gillian Nave of NIST
for clarification of a question regarding the \ion{Fe}{ii} spectrum.
We thank R.~Oudmaijer for comments on his reduction 
of the X-Shooter spectrum.
  
This research has used the SIMBAD database,
operated at CDS, Strasbourg, France. 
We made extensive use of the 
VALD\footnote{http://vald.astro.univie.ac.at/$\sim$vald/php/vald.php}
atomic data base \citep{kupka}, as well as the 
NIST\footnote{http://www.nist.gov/pml/data/index.cfm}
online Atomic
Spectroscopy Data Center \citep{Kram}.  Spectra were
obtained from the ESO Science Archive Facility under request number
SHUBRIG/2955, and CCOWLEY/101986.
  
\end{acknowledgements}


\rule{0mm}{5cm}

\begin{appendix}

\section{Data on iron group elements}\label{appendixB}
\begin{table}[ht!]
\small
\caption{Data for lines used for column densities
of elements other than iron.  References are to
the oscillator strength source.\label{tab:others}}
\centering
\begin{tabular}{l c c r} \hline \hline
{\Rv}Wavelength [\AA]&$\chi$\,(eV)&$\log(gf)$&$W_\lambda$\,(m\AA) \\ \hline
\multicolumn{4}{c}{{\Rv}\ion{Sc}{ii}, \citet{LawDak}} \\ \hline
{\Ru} 4246.82   & 0.315&   0.242   &  616  \\
 4400.49   & 0.605&  $-$0.536 &    304  \\
 4415.56   & 0.595&  $-$0.668 &    491  \\
 4420.67   & 0.618&  $-$2.273 &     20  \\ \hline
\multicolumn{4}{c}{{\Rv}\ion{Ti}{ii}, \citet[2002]{Picker}} \\ \hline 
 {\Ru}4025.13   & 0.61 &  $-$2.14 &     310  \\
 4028.34   & 1.89 &  $-$0.96 &     318  \\
 4227.35   & 1.13 &  $-$2.24 &     169  \\
 4287.87   & 1.08 &  $-$1.82 &     250  \\
 4290.22   & 1.17 &  $-$0.85 &     455  \\
 4300.05   & 1.18 &  $-$0.44 &     532  \\
 4312.86   & 1.18 &  $-$1.10 &     434   \\
 4391.03   & 1.23 &  $-$2.28 &     166  \\
 4398.29   & 1.22 &  $-$2.78 &      86  \\
 4399.77   & 1.24 &  $-$1.19 &     414  \\
 4411.93   & 1.22 &  $-$2.52 &      98  \\
 4432.11   & 1.24 &  $-$2.81 &      47  \\
 4441.73   & 1.18 &  $-$2.27 &     190  \\
 4443.79   & 1.08 &  $-$0.72 &     430  \\
 4450.48   & 1.08 &  $-$1.52 &     418  \\
 4464.45   & 1.16 &   1.81   &   276  \\
 4468.51   & 1.13 &  $-$0.60 &     506  \\
 4501.27   & 1.12 &  $-$0.77 &     568  \\
 4518.33   & 1.08 &  $-$2.91 &      91 \\ \hline
\multicolumn{4}{c}{{\Rv}\ion{Cr}{ii}, \citet{Nilsson}} \\  \hline	  
{\Ru}4195.42    & 5.32 &  $-$1.95  &     41  \\
4242.36    & 3.87 &  $-$1.17  &    324  \\
4558.66    & 4.07 &  $-$0.656 &    616  \\
4588.20    & 4.07 &  $-$0.627 &    545  \\
4634.07    & 4.07 &  $-$0.980 &    454  \\
4812.34    & 3.86 &  $-$1.800 &    341  \\
4824.13    & 3.87 &  $-$0.920 &    500  \\
4836.23    & 3.86 &  $-$2.25  &    137  \\   \hline
\multicolumn{4}{c}{{\Rv}\ion{Ni}{ii}, \citet{kur14}} \\ \hline
{\Ru} 4015.47   & 4.03 &  $-$2.41  &     58  \\
 4067.03   & 4.03 &  $-$1.83  &    269  \\
 4244.78   & 4.03 &  $-$3.10  &     17  \\
 4362.10   & 4.03 &  $-$2.70  &     59  \\ \hline
\end{tabular}
\end{table}

\end{appendix}

\end{document}